\documentclass[structabstract]{aa}  
\usepackage{graphicx}
\usepackage{txfonts}
\begin{document}
   \title{The peculiar nova V1309 Sco/Nova Sco 2008\thanks{Based on data collected at UT2+UVES within the program 282-D.5055(A/B).}}

   \subtitle{A candidate twin of V838 Mon}

   \author{Elena Mason
          \inst{1}
          \and
          Marcos Diaz\inst{2}
	  \and 
	  Robert E. Williams\inst{3}
	\and 
	George Preston\inst{4}
	\and 
	Thomas Bensby\inst{1}
          }

   \institute{ESO, Alonso de Cordova 3107, Vitacura, Santiago, CL\\
              \email{emason@eso.org,tbensby@eso.org}
         \and
             Departamento de Astronomia - Universidade de Sao Paulo, 05508-900 Sao Paulo, BR\\
             \email{marcos@astro.iag.usp.br}
	 \and 
	 STScI, Baltimore, MD, USA \\
	\email{wms@stsci.edu}
	\and 
	Carnegie Observatories, Pasadena, CA, USA
	\email{gpreston@ociw.edu}
             }

   \date{Received: Nov 2009; accepted: Apr 2010}

 
  \abstract
   {}  
   {Nova Scorpii 2008 was the target of our Directory Discretionary Time proposal at VLT+UVES in order to study the evolution, origin and abundances of the heavy-element absorption system recently discovered in 80\% of classical novae in outburst.} 
   {The early decline of Nova Scorpii 2008 was monitored with high resolution echelle spectroscopy at 5 different epochs. The analysis of the absorption and the emission lines show many unusual characteristics.}  
   {Nova Scorpii 2008 is confirmed to differ from a common Classical Nova as well as a Symbiotic Recurrent Nova, and it shows characteristics which are common to the so called, yet debated, red-novae. The origin of this new nova remains uncertain.}  
   {}


   \maketitle
%

\section{Introduction}

Nova Scorpii 2008 was independently discovered at mag $\sim$9.5-10 (unfiltered CCD) on Sep 2.5 UT by K. Nishiyama, F. Kabashima and Sakurai in  Japan and by Guoyou Sun and Xing Gao in China (CBET 1496). Nakano (2008) reports precise coordinates of the new object, noting that it was invisible on earlier observations taken by the same authors on Aug 20.5 and 21.5 UT (limit magnitude 12.8 and 12, respectively) and that is only 1.14 arcsec away from the USNO-B1.0 star 0592-0608962 of magnitude B=16.9 and R=14.8.
The first low resolution (R$\sim$500 and 1200) optical spectra were secured by Naito and Fujii (2008) who observed V1309 Sco on 3 consecutive days (Sep 3 to 5) reporting a smooth continuum, some absorption lines, and strong Balmer lines.
Rudy et al. (2008a) observed V1309 Sco in NIR spectroscopy on Sep 5 and confirmed it to be a slow nova at very early stages with some emission lines (early H Paschen series and FeII), but mostly absorption lines (H Paschen and Brackett as well as CaII, NI and CI). The lines were narrow with FWHM not exceeding 300 km/s. One month later the same group obtained additional NIR spectroscopy between 1-5 microns that showed a strong continuum resembling that of a late M giant star, upon which were superposed strong molecular absorption from CO, H\_2O, and weaker features of TiO and VO (see Rudy et al. 2008b).

Recently, Williams et al. (2008) have shown that the majority of classical novae (CNe) in outburst show a short lived absorption system from heavy element (Transient Heavy Element Absorption -THEA- system) which is external to the primary expanding ejecta. The numerous early reports of absorption lines in the spectra of V1309 Sco motivated us to apply for Director’s Discretionary Time (DDT) on the VLT+UVES to better characterize the THEA system in classical novae. 

In this paper we present the spectra we obtained and their analysis.


\section{Observations and data reduction}

V1309 Sco was observed during five epochs in the interval [+10,+47] days following discovery until it disappeared   behind the Sun. The instrument setup was the same for each epoch and the exposure times were progressively increased assuming a typical classical nova light curve and line intensities \footnote{The program was scheduled as a  time critical one and was not a target of opportunity program.}. This resulted in the saturation of the H$\alpha$ emission lines in the epoch 5 spectrum, due to the anomalous Balmer lines intensity developed by V1309~Sco. 
At each epoch we were observing with the UVES standard dichroic setups, namely DIC1 346+580 and DIC2 437+860. The combination of the two dichroics mode of observations allows coverage of the spectral range $\sim$3000-11000\AA, with only two gaps between 5757-5833 and 8520-8658~\AA. 
The detailed log of the observation is reported in Table~\ref{tab1}.  

Calibration frames (arc, flats, order definition etc) were taken during the morning following the observations according to the UVES calibration plan. The data were reduced using the UVES pipeline v3.6.8 and the optimal extraction mode. 
The observation of the spectrophotometric standard EG~21 at the time of epoch~3 spectrum allowed us to remove the instrument signatures and perform relative flux calibration of our data assessing the continuum shape (SED). We did not attempt to determine broad band magnitudes via convolution of the spectra with broad band filter transmission functions, because we could not estimate the slit losses.  

Medium resolution spectra of V1309 Sco were also obtained at the SOAR~4.1~m telescope as part of an ongoing ToO program aimed at monitoring the spectral evolution of classical novae in outburst. The Goodman Spectrograph was employed to take spectra from 350 to 900 nm with a spectral resolution R$\sim$4000 (0.16 nm FWHM resolution at H$\alpha$). A narrow slit of 0.46'' was used to reach the maximum possible resolution with a 600 gr/mm Volume Phase Holographic (VPH) grating. Spectrophotometric flux calibration of most of the spectra was achieved by using wide-slit exposures and standard stars (Hamuy et al. 1994) observations. Standard reduction procedures including the optimal extraction of faint spectra 
were applied to the data. A total of 7 different epochs were sampled from $\sim$1 day after maximum to late April, 2009. In June 2009 the object was found too faint for continuing the SOAR observations. 

In addition, a spectrum of the nova was observed on September~11 2009 with the high resolution (R$\geq$53000 and $\geq$42000 in the blue and red camera, respectively) echelle spectrograph MIKE (Bernstein et al. 2003) at the 6.5m-Magellan telescope in Las~Campanas. The blue camera covers the wavelength range 320-500 nm, while the red camera covers the wavelength range 490-1000 nm.  Several spectra with different exposure times were taken with each camera in order to avoid possible saturation of the strongest emission lines. The data was reduced with the IDL pipeline\footnote{Available at http://web.mit.edu/?burles/www/MIKE/} developed by Burles, Prochaska, and Bernstein. No flux calibration nor order merging were performed. 

\begin{figure}
   \centering
   \includegraphics[width=9cm]{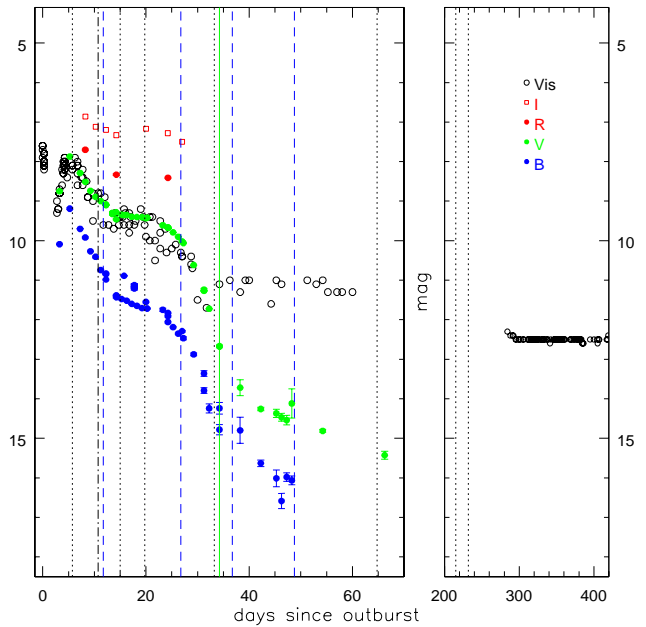}
   \caption{V1309 Sco light curve observed in Vis, B, V, R and I band by the AAVSO team (pre-validated data points). The different bands are plotted in different colors and symbols (see  the color code in the figure itself). The vertical lines mark the epochs of our observations: blue dashed lines are for UVES observations, black dotted lines are for the Goodman observations and the black dashed-dotted lines is for the MIKE one.}
              \label{fig0}%
    \end{figure}

The log of SOAR and Magellan observations are reported in Table~\ref{tab1}, too. 
In addition, in Fig.~\ref{fig0} we plot the AAVSO light curve together with the epochs of our spectroscopic observations (marked by vertical lines). 

In this paper we focus our analysis on the VLT+UVES observations because of their higher quality. We used the MIKE and SOAR spectra to derive complementary information as possible. 

The data analysis was performed on dereddened spectra. We estimated the reddening following the empirical law determined by Munari and Zwitter (1997) and measuring the EW of the KI~$\lambda$7699.0 interstellar absorption line in our MIKE and UVES spectra. We computed $<$EW$>\simeq$0.14\AA \ by averaging the measurement from 18 different spectra and derived E(B-V)$\simeq$0.55 mag. We then dereddened the spectra assuming R=3.1.

\begin{table*}
\caption{Log of the observations.}
\label{tab1}
\centering
\renewcommand{\footnoterule}{}  
\begin{tabular}{cccccccc}
\hline \hline
spectrum ID & UT date & exp-time (s) & tel+inst & setup & slit width ('') & seeing ('') & sky transparency \\
\hline
SOAR a & Sep 07.03 2008 & 30(B),5(R) & SOAR+Goodman & - & 0.46 & 1.3 & CLR \\
MIKE blue   & Sep 11.99 2008 & 5$\div$400  & Magellan Mike & blue &  0.7  & 0.9 & ?  \\
MIKE red    & Sep 11.99 2008 & 5$\div$400  & Magellan Mike & red  &  0.7  & 0.9 & ?  \\
epoch~1 & Sep 13.02 2008 & 150$\times$3 & UT2+UVES & DIC1 346+580 & 0.5 & 1-1.2 & CLR \\
epoch~1 & Sep 13.03 2008 & 150$\times$3 & UT2+UVES & DIC2 437+860 & 0.5 & 1-1.2 & CLR \\
SOAR b & Sep 16.03 2008 & 240(B),60(R) & SOAR+Goodman & - & 0.46 & 0.9 & CLR\\
epoch~2 & Sep 20.04 2008 & 200$\times$3 & UT2+UVES & DIC2 437+860 & 0.5 & 1.25-2.25 & THN \\
epoch~2 & Sep 20.98 2008 & 200$\times$3 & UT2+UVES & DIC1 346+580 & 0.5 & 1.25-2.25 & THN \\
SOAR c & Sep 21.01 2008 & 60 & SOAR+Goodman & - & 0.46 & 0.9 & CLR \\
epoch~3 & Sep 28.02 2008 & 300$\times$2 & UT2+UVES & DIC1 346+580 & 0.5 & 0.6-0.8 & CLR \\
epoch~3 & Sep 28.03 2008 & 300$\times$2 & UT2+UVES & DIC2 437+860 & 0.5 & 0.6-0.8 & CLR \\
SOAR d & Oct 04.04 2008 & 120 & SOAR+Goodman & - & 0.46 & 1.3 & CLR \\
epoch~4 & Oct 08.04 2008 & 600$\times$2+120 & UT2+UVES & DIC1 346+580 & 0.9(B), 0.7(R) & 0.50-0.65 & CLR \\
epoch~4 & Oct 08.06 2008 & 600$\times$2+120 & UT2+UVES & DIC2 437+860 & 0.0(B), 0.7(R) & 0.50-0.65 & CLR \\
epoch~5 & Oct 20.01 2008 & 600$\times$3 & UT2+UVES & DIC1 346+580 & 0.9(B), 0.7(R) & 0.50-0.65 & CLR \\
epoch~5 & Oct 20.04 2008 & 600$\times$3 & UT2+UVES & DIC2 437+860 & 0.9(B), 0.7(R) & 0.50-0.65 & CLR \\
SOAR e & Nov 05.03 2008 & 600(B),200(R) & SOAR+Goodman & - & 0.46 & 1.4 & THN \\
SOAR f & Apr 04.37 2009 & 500(R) & SOAR+Goodman & - & 0.46 & 1.0 & CLR \\
SOAR g & Apr 21.33 2009 & 900(B),600(R) & SOAR+Goodman & - & 0.46 & 1.0 & CLR \\
\hline
\end{tabular}
\end{table*}

   \begin{figure*}
   \centering
   \includegraphics[width=18cm]{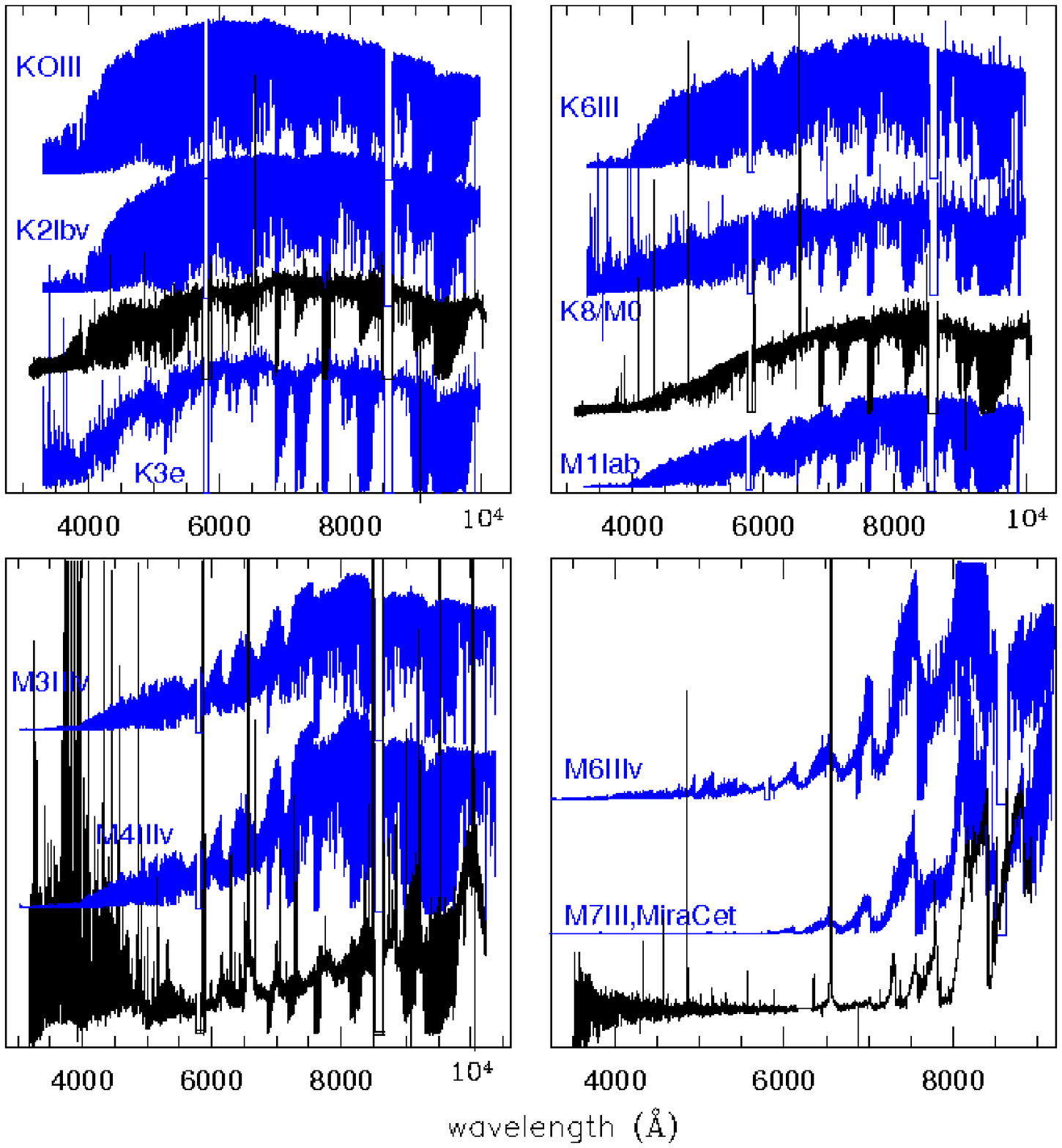}
   \caption{The comparison of V1309 Sco spectra with those of known bright stars from the UVES-POP database. Upper left panel: epoch~1; upper right panel: epoch~3; lower left panel: epoch~5; lower right panel: SOAR April 20 spectrum. V1309~Spectra have been dereddened as explained in Section~2. The UVES-POP spectra have been arbitrarily scaled in flux for an easier comparison. They have not been dereddened.}
              \label{fig1}%
    \end{figure*}

\section{Early spectral evolution}

The early evolution of V1309~Sco was characterized by significant  changes in both the continuum and the lines. 
During the period of our spectroscopic observations the continuum changed from peaking in the visible range during the early epochs, where its max intensity shifted from $\sim$6500-7000 \AA \ at epoch 1 to $\sim$7500-8000 \AA \ at epoch 3, to become much flatter and cooler at later epochs, possibly peaking in the NIR/IR. In none of our spectra did we observe a blue continuum as typically observed in the case of classical novae outbursts. 

In order to better understand the object SED we searched the UVES-POP (UVES Paranal Observatory Project\footnote{http://www.sc.eso.org/santiago/uvespop/interface.html}) data base for high resolution spectra of cool stars to find spectra with similar characteristics that would enable us to compare the spectra of V1309 Sco with known objects. 
Fig.~\ref{fig1} shows stellar spectra (blue lines) that most closely match each of our V1309~Sco spectra (black line). The comparison shows that in little more than 1 month V1309 Sco has evolved from and early K-type giant (upper left panel of Fig.2) to a late K - early M type giant (upper right panel of Fig.2), to possibly late M type star (lower left panel). The temperature and the SED of our target appeared to cool from $\sim$ 5-6000K to 4000K and even  cooler. The SOAR observations of April 2009 (lower right panel of Fig.2) confirm that V1309 Sco has continued its evolution to lower temperatures, with its spectrum resembling that of a M6-M7 giant after 213-229 days since outburst.

   \begin{figure}
   \centering
   \includegraphics[width=9cm]{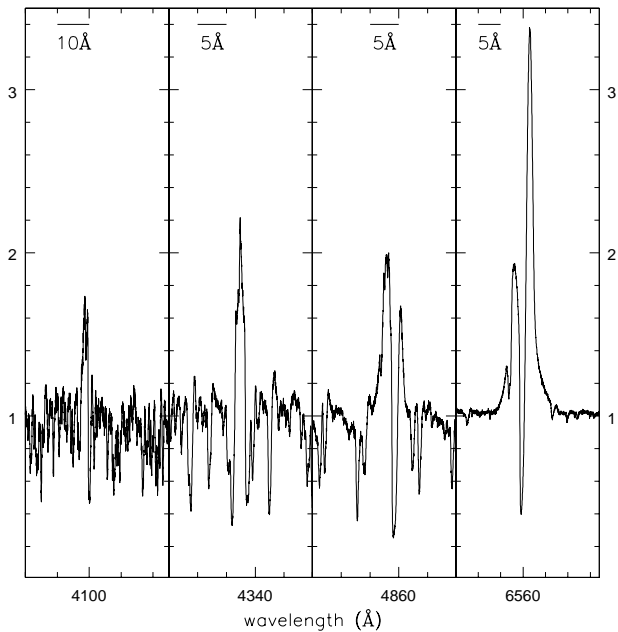}
\caption{H$\alpha$, H$\beta$, H$\gamma$ and H$\delta$ line profile in V1309~Sco epoch~1 spectrum. The lines blueward of H$\zeta$ are in absorption. The different profiles imply that the region where the lines form is not optically thin and probably not homogeneous either. The epoch~1 spectrum has been normalized before plotting.  }
              \label{fig2}%
    \end{figure}

\subsubsection{The Balmer lines}

The evolution of the lines is equally complex and interesting. 
The spectra are dominated by absorption lines during the first three epochs and by emission lines during the last two. The strongest emission lines at all epochs are those from the Balmer series of hydrogen. 

The Balmer lines show very complex profiles and evolution. In the epoch 1 spectrum the higher lines of the series are in absorption and only the early lines of the series show emission components. In addition, those lines which are in emission  
show a distinct profile (Fig.~\ref{fig2}). By the time of epoch 3 the Balmer series is in emission up to H21. The Balmer series line intensity increases by a factor $\sim$20-25 in the epoch 4 and 5 spectra, while the continuum emission is weaker and redder. 

The Balmer emission lines at early epochs (epoch 1 to 3) can all be fitted by a double Gaussian, namely by a narrow absorption line superposed to a stronger and broader Gaussian emission line. The emission components have average FWHM (H$\alpha$ and H$\beta$) of $\sim$150 km/s. The Gaussian absorption components have narrower FWHM ($\sim$ 80 km/s) and their velocity relative to the emission component is $+$15 km/s, i.e., the emission component is more blue shifted than the absorption component\footnote{The only exception is the H$\alpha$ line in the epoch 1 spectrum as shown in Fig.~\ref{fig2}.}. The Balmer emission lines at later epochs are better fit by a double Gaussian emission line, with a strong blue peak flanked by a weaker red peak. The ''minimum'' between the two Gaussian emission components perfectly matches in position the absorption component of the previous epochs. 
The profiles have some similarity to inverse P-Cyg profiles, although we believe that the line profiles are much closer to those from an axis-symmetric shell or a collimated wind.  Inverse P-Cyg profiles imply  in-falling material onto the stellar surface. However, in our spectra the absorption lines, although red-shifted with respect to the emission components, are all blue-shifted with respect to the system velocity (see below) implying that the gas which is responsible for the absorption is moving away from the central object. 
The absorption lines superposed on the emission, together with the broad extended wings (see text below) compose a profile which is somehow similar to that seen in pre-PN (PPN) and young PN or post AGB stars (e.g. Sanchez-Contreras et al. 2008).

   \begin{figure}
   \centering
   \includegraphics[width=9cm]{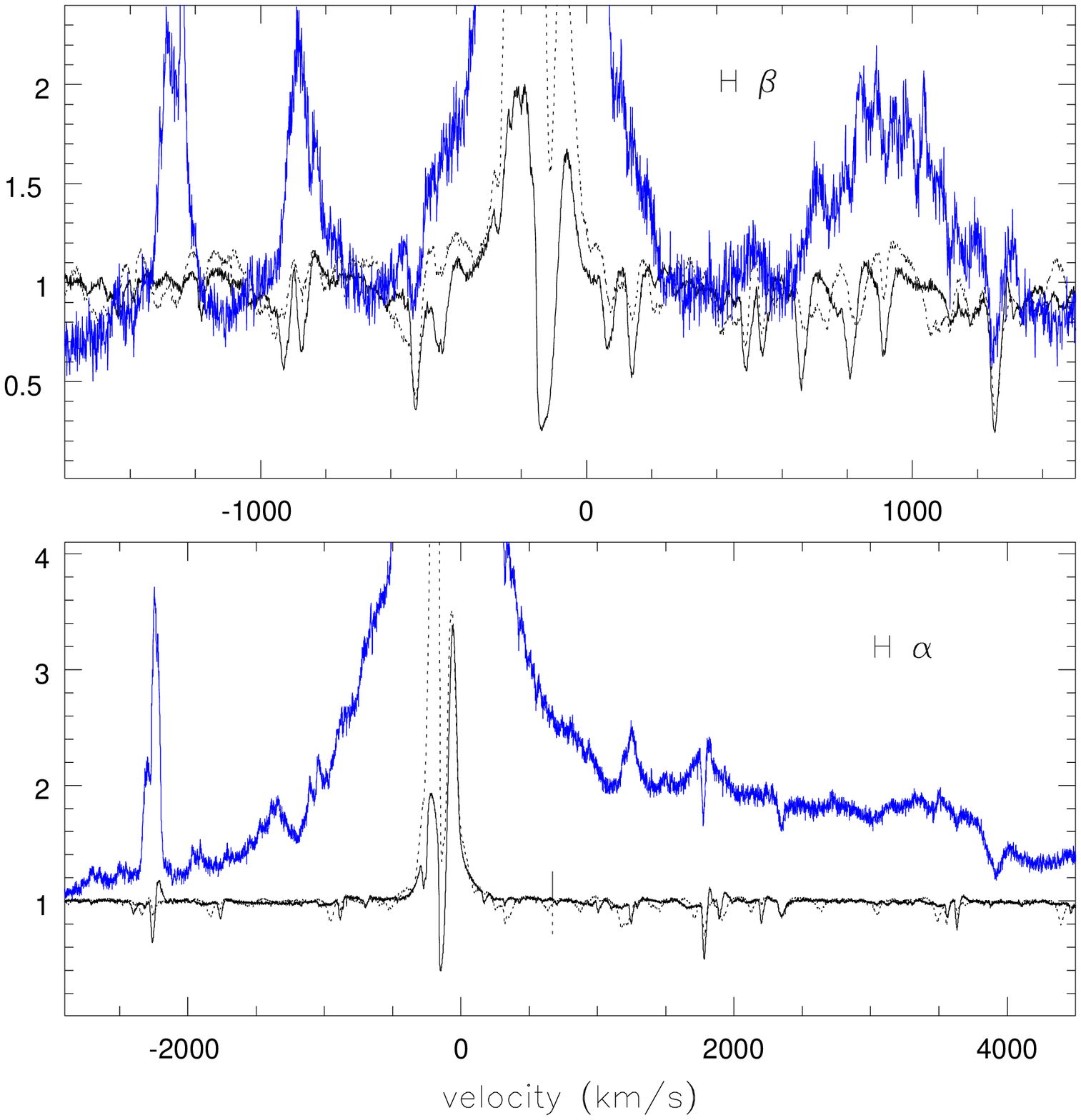}
   \caption{The wings of the H$\alpha$ and H$\beta$ emission lines in the velocity space.
The spectra have been corrected for the systemic velocity =-80 km/s
(see text) and are normalized (y axis is in relative flux units). The black
solid line is for the epoch 1 spectrum; the black dotted line is for epoch
3 spectrum and the blue solid line is for the epoch 5 spectrum. Epoch
2 spectrum is very much similar to epoch 1 and 3 spectra; while epoch
4 spectrum is very similar to the epoch 5 one and are not shown in
the figure for clarity. The extended wings of the Balmer lines appear
with epoch 4 and 5 spectra. The wings of the H$\beta$ emission line are less
extended than the H$\alpha$ ones by a factor 2-3.}
              \label{fig3}%
    \end{figure}

\begin{figure}
   \centering
   \includegraphics[width=9cm]{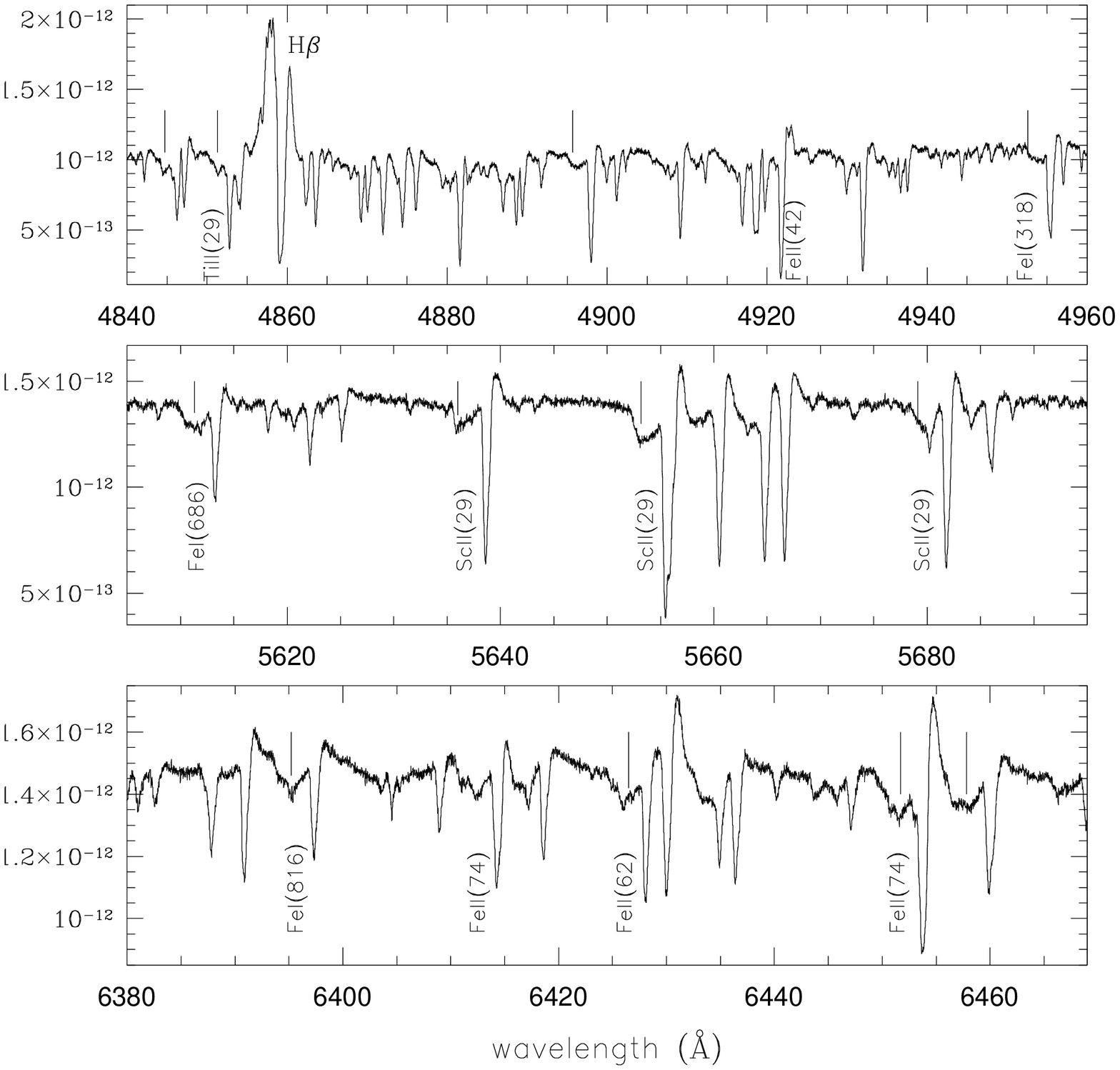}
   \caption{An example of the s-element P-Cyg profile and absorption
troughs at the time of epoch 1 spectrum. The three panels in the figure
have flux units of erg~s$^{-1}$~cm$^{-2}$~\AA$^{-1}$ along the y-axis. The vertical lines
mark some of the most evident absorption troughs, while a few line ID
have been reported on the side of the line absorption component.}
              \label{fig4}%
    \end{figure}

Close inspection reveals that the H$\alpha$ and H$\beta$ Balmer lines show minor absorption and emission components as well as extended wings. The H$\alpha$ wings extend up to velocities of 320 km/s in early epochs and up to $\sim$2000-3500 km/s in the epoch 4 and 5 spectra (Fig.~\ref{fig3}). The wings of the H$\beta$ and the other lines of the Balmer series are never as extended as those of the H$\alpha$ emission line and the high velocity material reaches velocities up to 300-500 km/sec only. The H$\beta$ lines shows an extended trough which survives during the first 3 epochs. This is observed also in some of the strongest THEA  absorption lines in the epoch 1 spectrum (see text below and Fig.~\ref{fig4}).

\subsubsection{Heavy elements absorption lines}

The absorption lines which dominate the early epochs spectra are from neutral or singly ionized heavy elements, especially the Fe-peak elements.  
In particular, we have identified lines from the following species (multiplets): Ti II (multiplets 12, 11, 87, 21, 20, 41, 94, 51, 93, 31, 18, 30, 50, 60, 38, 48, 92, 29, 70, and others), Fe II (multiplets 172, 39, 28, 27, 20, 32, 37, 38, 42, 43, 25, 49, 55), Fe I (multiplets 43, 3, 42, 15, 152, 41, 68, 2, 15, 62, 318, 686 and 816), Sc II (multiplets 15, 14, 29 and 31), V II (multiplets 9, 32, 25 and 37), Cr II (multiplets 19, 31, 44, 30, 43, 23), Cr I (1), Sr II (1), BaII (2), Y II (5 and 27), Ca I (2 3 and 21), Mn I (1), Mg I (2, 7, 8 and 9), Na I~D and (6), etc. 
The strongest lines, e.g. Fe II 40, 74, 49, 42, Sc II 29 and 31, as well Ti II, show a P-Cyg like profile having a deep narrow absorption embedded in a much weaker (by a factor 2 or 4) emission feature which often has a relatively broad red wing.  As in the case of the H$\beta$ line, these heavy element and s-process P-Cyg lines also show an absorption trough on the blue side of the P-Cyg absorption (see, e.g., Fig.~\ref{fig4}). 
We measured for all the heavy element absorption lines (with or without P-Cyg profile) an average FWHM of $\sim$30 km/s, which is the narrowest for any THEA system so far observed. We determined for all of them a velocity of about -50 km/s with respect to the systemic velocity. The binary system radial velocity was estimated from the [CaII] $\lambda\lambda$7291,7321 doublet which was visible at all epochs. The average radial velocity computed from the Ca II frobidden lines in the 5 epochs is -80$\pm$15 km/sec. 
The absorption troughs have an average velocity of +150$\pm$18 km/s with respect to the above systemic velocity.

   \begin{figure*}
   \centering
   \includegraphics[width=18cm]{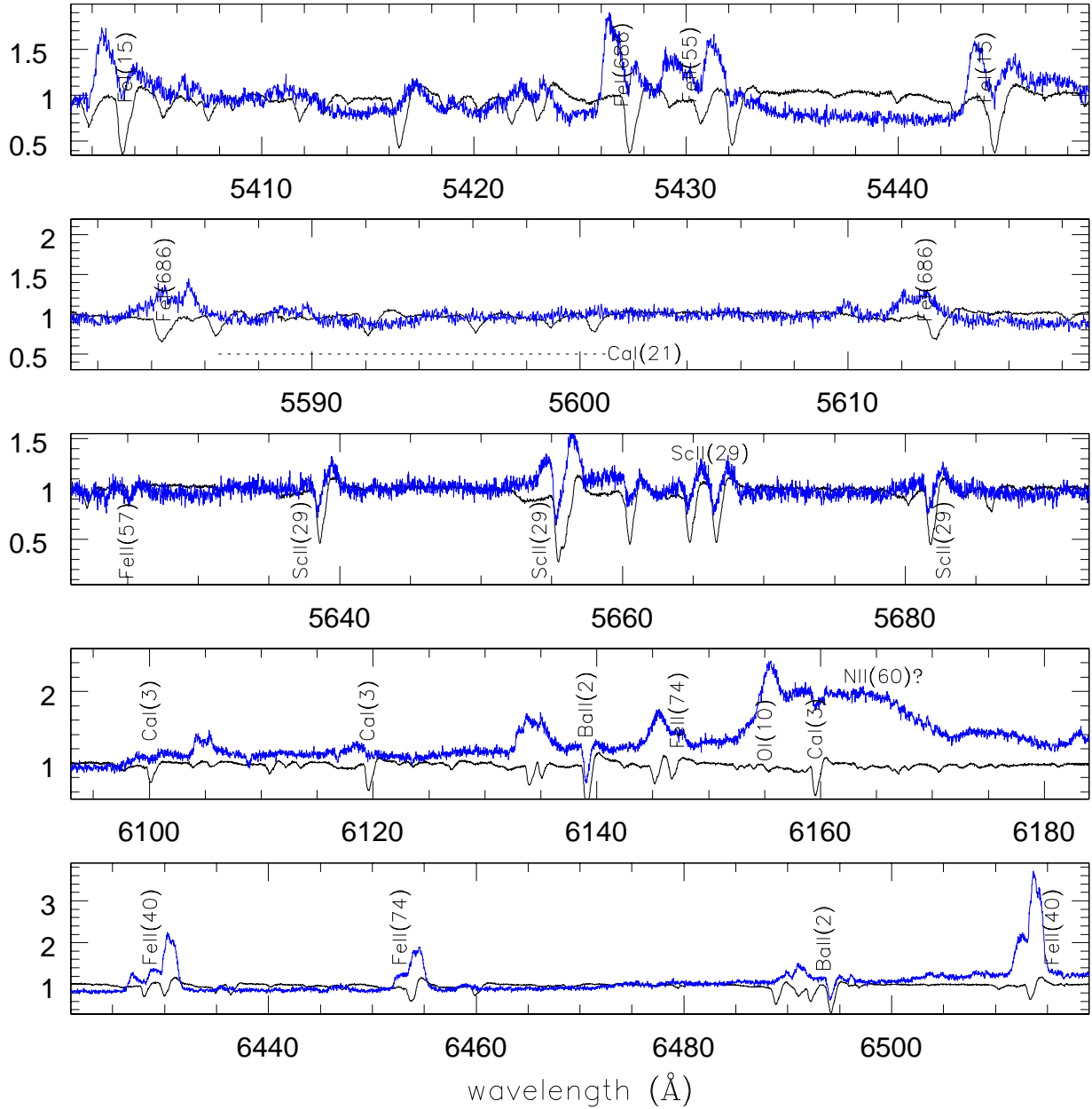}
   \caption{Example of the spectral evolution of the s-elements absorption lines into composite emission line profile. The black solid line is for the epoch 1 spectrum, while the blue solid line is for the epoch 5 spectrum. The two spectra have been normalized and the y axis are in relative flux units.}
              \label{fig5}%
    \end{figure*}

The evolution of the THEA absorption and P-Cyg lines is rather unusual. 
Figure~\ref{fig5} show examples of the evolution of different elements, comparing profiles from epoch1 (black solid line) and epoch~5 (blue solid line) spectra. As an example, the CaI (21) multiplet is in absorption at epoch~1, but has disappeared by epoch~5. On the other hand, CaI(6) has evolved into a blue shifted emission peak at epoch~5. The BaII (2) deep absorptions survive until epoch~5, developing only weak emission wings at that time. The ScII (29) absorption is deep and survives during the epochs of our observations, but develops a stronger emission component having the redshifted peak stronger than the blue shifted peak. The FeII (46, 40 and 74) absorption lines evolve into emission lines: the absorption is greatly reduced and superposed to an asymmetric emission lines characterized by a red component which is stronger than the blue one. On the contrary, FeI (15) develops a stronger blue emission component. FeII (42) evolves similarly to FeI  (15). 
In general, the spectral evolution of the THEA absorptions is toward development of associated emission components. The asymmetry and the differences in the emission profiles can probably be ascribed to an inhomogeneous line forming region.  

We should note the presence of HeI (11), (18), (46) and (10) in epoch~1 spectrum. More HeI (multiplets 2, 14, 48, 4, 10  transitions and other emission lines from higher potential energy elements such as OI (1\footnote{However the identification is uncertain as the triplet does not show emission of roughly equal intensity but a very strong line at $\lambda$7772 and much weaker emission at $\lambda$7774 and $\lambda$7775. 
}, 4, and 10 and possibly 14), NII (60, though it has to be confirmed, as no other NII transitions have been detected in the spectra) and HeII(1)~$\lambda$4686 become visible in the epoch 4 and 5 spectra. These late spectra show also evidence of the forbidden transition [OI](1)~$\lambda\lambda$6300,6364 and (3)~$\lambda$5577. 

   \begin{figure}
   \centering
   \includegraphics[width=9cm]{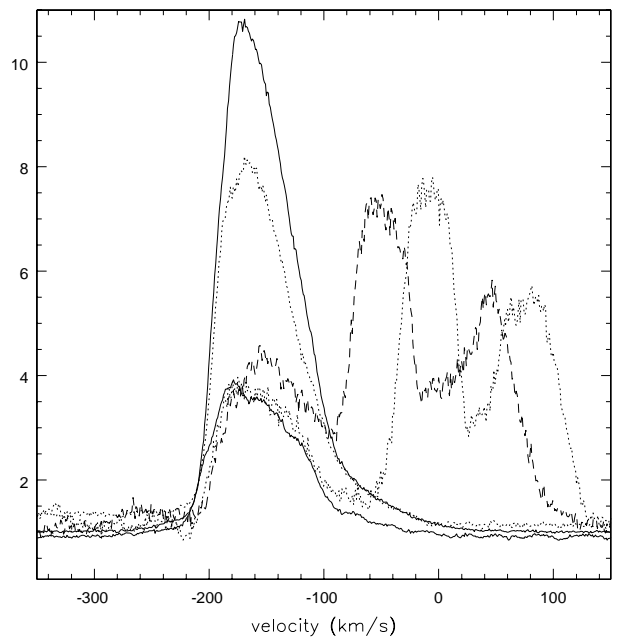}
   \caption{The line profile of the HeI (11, 10, 46, 4 and 48) emissions as in our epoch~5 spectrum. Different line style are just for clarity The spectrum has been
normalized to the continuum and the line intensity of HeI (11) has been arbitrarily scaled for clarity. The HeI lines are all centered at $v\sim$160-180 km s$^{-1}$ with respect to the observer. The line velocity in the reference system of V1309~Sco is reduced by 80 km s$^{-1}$.}
              \label{fig6}%
    \end{figure}

\begin{figure}
   \centering
   \includegraphics[width=9cm]{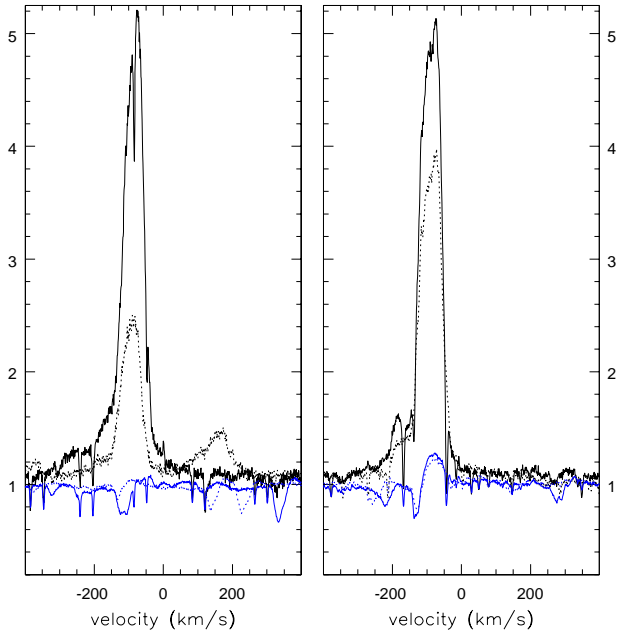}
   \caption{The line profile of the [OI](1) doublet (left panel) and the
[CaI](1) doublet (right panel) in epoch 5 spectrum. The black solid line
is for the strongest emission in each doublet, while the blue solid line is
of the weaker emission lines. The black dashed line shows the strongest
line of each doublet in epoch 1 spectrum. The red-dashed lines shows
the weak line of each doublet in epoch 1 (the CaI line profiles in epoch
1 are remarkably similar including the absorption. Spectra are in the velocity
space in the system frame of reference and flux are relative to the
normalized continuum.}
              \label{fig7}%
    \end{figure}

These higher excitation potential emission lines have a different profile than other lines. In particular, the HeI emission lines are characterized by an asymmetric Gaussian profile with a steep blue wing and a very extended red wing (see Fig.~\ref{fig6}). This type of profile has been observed in PPN and AGB stars, too (see e.g. Sanchez-Contreras et al. 2008). 
The forbidden lines are a-symmetric showing a more extended blue wing (Fig.~\ref{fig7}). This blue wing is sufficiently weak that it can be ignored in our use of the forbidden lines for the computation of the systemic velocity.

In April 2009 SOAR spectra the strongest lines belong to the permitted transition of the Balmer series, FeII~(42), (49) and (55) multiplets and the CaII~(2) triplet and the forbidden [OI] (multiplet 1 and 3) and [CaII](2) transitions already detected in the early spectra. We could also observe few other new and emission lines which we tentatively identify with [FeII](19)~$\lambda$5261 and [FeII](18)~$\lambda$5155 and $\lambda$5273, though we are unable to the detect all the transition belonging to those multiplets because of the low SNR of the spectrum. 

In addition, the April SOAR red spectra shows molecular features at $\lambda$7050, 7589, 8454 and 8862 \AA, which we identify with TiO band heads (Rayner et al. 2009, Kaminski et al. 2009). While, we identify with VO band heads the features at $\sim$7330  and 7830 \AA \ (Martini et al. 1999). Following the same analysis done by Martini et al. (1999) on V4332~Sgr spectra, we conclude that the April spectrum of V1309~Sco resembles that of a M8-9 giant. 

\section{Interpretation/analysis}

\begin{table*}
\caption{V1309 Sco Balmer decrement observed at different epochs and compared with theoretical predictions (see text for more details). The line ratios were measured on dereddened spectra assuming E(B-V)=0.55 mag. }
\label{tab2}
\centering
\renewcommand{\footnoterule}{}  
\begin{tabular}{ccccccccc}
\hline \hline
Line ratio & epoch~1 & epoch~3 & epoch~4 & epoch~5 & SOAR-e & SOAR-g & case-B & accretion disk \\
\hline
H$\alpha$/H$\beta$      & 5.15 & 2.86 & 2.17 & -    & $\sim$14$^\dagger$ & $\sim$50$^\dagger$ & 3.05 & 1.23 \\
H$\beta$/H$\beta$       & 1.00 & 1.00 & 1.00 & 1.00 &   1.00 & 1.00       & 1.00 & 1.00 \\
H$\gamma$/H$\beta$      & 0.74 & 0.60 & 1.20 & 0.96 &      0.28    & 0.20 & 0.45 & 0.75 \\
H$\delta$/H$\beta$      & 0.44 & 0.33 & 0.59 & 0.52 &      0.12    & -   & 0.25 & 0.58 \\
H$\varepsilon$/H$\beta$ & -    & -    & -    &  -    &       - & -        & 0.15 & 0.46 \\
H$_8$/H$\beta$          & 0.31 & 0.12 & 0.23 & 0.18 &           - & -    & 0.10 & 0.38 \\
H$_9$/H$\beta$          & -    & 0.09 & 0.17 & 0.15 &           - & -    & 0.07 & 0.32 \\
H$_{10}$/H$\beta$       & -    & 0.05 & 0.12 & 0.10 &           - & -    & 0.05 & 0.27 \\

\hline
\end{tabular}
\scriptsize
\begin{center}
$^\dagger$ The H$\alpha$/H$\beta$ ratio is only approximate due to relative calibration errors between the blue and red spectra taken at SOAR. 
\end{center}
\end{table*}

The line profiles observed in the late epochs spectra and the strongest emission lines observed for all epochs, are consistent with the line forming region being an optically thick slow wind which ejects material either in subsequent multiple layers or in an axis-symmetric partially collimated outflow, similarly to the case of PPN and PAGB stars (Sanchez-Contreras et al. 2008). 
In particular, we can explain the observation of absorption lines superimposed on broader emission by a slowly expanding shell which is denser in its equatorial plane. The dense equatorial belt produces the absorption lines; while the rest of the shell, being less dense, evolves into an emission line region. The expanding shell probably resemble an oblate spheroid inclined with respect to the line of sight in a way that we mostly see just the polar cap/hemisphere which is approaching the observer. This would explain the blueshift of both the absorption and the emission line components.The ejecta from the polar caps is also characterized by a wide range of expansion velocities, thus explaining the high velocity extended wings.
In addition, the expanding shell is not symmetric and homogeneous as shown by the different line profiles we observed for different elements and different multiplets from a same element. The later development of emission lines from higher ionization potential elements (e.g. HeI and OI) and forbidden emission ([OI]), both with yet different profiles, is further evidence of the broad range of physical conditions within the gas which  produces the V1309~Sco spectrum.

We have measured several line flux ratios in the tentative to better characterize these gas regions.
We report in Table~\ref{tab2} the Balmer decrement measured from H$\alpha$ to H10 at all epochs but the second one. We compare those with the Balmer decrement predicted for optically thin case-B regions (Osterbrock 1989) 
and optically thick gas as expected in accretion disks (Williams 1980). In no epoch our observed Balmer decrement is close to the optically thin case-B of recombination or as flat as in the case of Planckian optically thick lines.  The Balmer decrement blueward of H$\beta$ is always steeper than predicted by theory, impling that collisions and self-absorptions are playing an important role in the line formation mechanisms (Osterbrock et al. 1976). In addition the H$\alpha$/H$\beta$ flux ratio which initially seemed to decrease (from epoch 1 to 4) has evolved toward large values at late epochs. The H$\alpha$ line was saturated in the epoch~5 UVES spectrum, while we measure line ratios as large as $\sim$14 and 50 in the SOAR spectra of November 2008 and April 2009, respectively. Steep  H$\alpha$/H$\beta$ Balmer decrement have been observed in symbiotic stars (mostly D-type and hosting a Mira donor star, Acker et al. 1998, Corradi et al. 2009) and explained with increased optical thickness, increased collisional excitation rate and presence of dust within the expanding nebula (Acker et al. 1998 and reference therein). 
Other peculiarities of the object Balmer decrement are: 1) the intensity of the H$\gamma$ line at epoch~4, which is stronger than the H$\beta$ emission and 2) the almost absolute lack of the Balmer line H$\varepsilon$~$\lambda$3970. We are unable to explain the ''H$\gamma$ excess'' with a line blend as the possible candidates HeII (3), OII (77) and SIII(4) do not show any significant emissions at the wavelength corresponding to the other transitions of the same multiplet. We explain the missing H$\varepsilon$ emission line with an absorption phenomena. The H$\varepsilon$ overlap in wavelength to the CaII~H line. The CaII~H-K doublet was observed to be in absorption and saturated at the early epochs (1 and 2) and to evolve toward emission starting from epoch~3. A high optical depth of the Ca~H line in an intervening gas cloud account for the undetected H$\varepsilon$. This is further confirmed by the considerable strength of the CaII(2) NIR triplet and the [CaII](1) doublet. By absorbing the 3934\AA \ photons the electrons populate the 4p~2P$_{3/2}$ level from which they decay producing $\lambda$8662, $\lambda$8542 and $\lambda$8498 CaII(2) and, by cascade down, the [CaII](1) emission lines. As a matter of fact we observe unusually strong CaII(2) and [CaII](1) emission lines if compared with other astrophysical objects (e.g. AGN, Ferland and Persson 1989; or T~Tau and Herbig star, Hamann 1994).  

We also measured the line ratio of the observed forbidden emission lines, namely [OI](1) and (3) and [CaII](1) which are the only forbidden transitions clearly detectable in our UVES (epoch~4 and 5) spectra. 
The [CaII](1) doublet is visible at all epochs but variable, decreasing in intensity at the time of epoch~3 spectrum and increasing afterward. The two [CaII](1) lines are relatively strong for accurate flux measurement in epoch~4 and 5 spectra which deliver the average line ratio CaII~$\lambda$7292/$\lambda$7324$\sim$1.33. This value is larger than the predicted value of 1 (Osterbrock 1951) implying that the lines are optically thick, consistently with the cascade-down process illustrated above. The [OI](1) doublet appears in the spectra of V1309~Sco starting from epoch~4. The average flux ratio which we measure in epoch~4 and 5 spectra is only slightly larger than the predicted value for the optically thin case, being it 3.1. This value is rarely observed in the classical novae ejecta (Williams 1994) which typically display [OI] flux ratios in the range 1-2 and which most likely form those emission lines in dense gas bubbles embedded within the expanding shell (Williams 1994). 

[CaII](1) emission lines have been observed in AGN (Ferland and Persson 1989) and in stellar objects such as T~Tau star and Ae and Be Herbig objects (e.g. Hamann 1994). Ferland and Persson (1989) use the relative intensity of the CaII forbidden and permitted emission lines, in combination with the Balmer ones, to constraint the physical condition of the gas. 
They note that small values of the ratio ($\lambda$7292+$\lambda$7324)/($\lambda$8494+$\lambda$8542+$\lambda$8662) are obtained in relatively dense environments and that such values decrease for increasing densities, once the collisional de-excitation mechanism became more efficient than the spontaneous radiative decay. The detection of the [CaII](1) lines implies that the electron density cannot be larger than N$_e\simeq$10$^{10}$~cm$^{-3}$. Ferland and Persson (1989) show that the ($\lambda$7292+$\lambda$7324)/($\lambda$8494+$\lambda$8542+$\lambda$8662) flux ratios measured in AGN are consistent with electron densities N$_e\geq$10$^9$~cm$^{-3}$, independently on the temperature T$_e$. 
Though we observe similar line ratios (we measure ($\lambda$7292+$\lambda$7324)/($\lambda$8494+$\lambda$8542+$\lambda$8662)$\sim$0.04-0.05\footnote{this value has been derived assuming that the CaII(2) are all of equal intensity as observed in the AGN and optically thick cases. The UVES spectra do not cover the two reddest lines of the multiplet.} in epoch~4, 5 and November 2008 SOAR spectra), we cannot straightforwardly conclude similar electron densities because we do not measure consistent flux ratios of the CaII NIR triplet over the H~Balmer or OI$\lambda$8446 lines but obtain, instead, significantly larger values\footnote{As an example, we measure ($\lambda$8494+$\lambda$8542+$\lambda$8662)/H$\alpha\sim$0.47-0.28 instead of 0.013-0.11; ($\lambda$8494+$\lambda$8542+$\lambda$8662)/H$\beta\sim$0.75-3.92 in place of 0.04-0.34 and ($\lambda$8494+$\lambda$8542+$\lambda$8662)/OI$\sim$20-27 in place of 0.35-3.24.}. Our anomalous high flux ratios are possibly all consistent with the pumping up mechanism described above. 

Hamann (1994) CLOUDY models describe the flux ratios [CaII]$\lambda$7921/[OI]$\lambda$6300 and [OI]$\lambda$5577/[OI]$\lambda$6300 as function of the electron density N$_e$. We measured average values of 1.59 and 0.16 for the two ratios, respectively, which, in the hypothesis of coronal ionization equilibrium, are consistent with the electron temperature T$_e\sim$18000~K and density N$_e\sim$10$^6$~cm$^{-3}$. 

This assumption too might be incorrect in the case of V1309~Sco as our measured [CaII] flux ratio does not  match the optically thin assumption and we do not observe other forbidden transitions as in the case of T~Tau and Ae \& Be Herbig stars. However, the [OI](1) flux ratio is consistent with optically thin conditions. Using the [OI] flux ratio ($\lambda$6300$+\lambda$6364)$\lambda$5577 to constraint the electron temperature in the hypothesis of optically thin gas, we find a relatively close matching solution: T$_e\sim$14-17000~K and N$_e\sim$10$^6$ cm$^{-3}$. Density larger than the critical density N$_e\sim$7$\times$10$^6$ should be discarded as otherwise the [CaII](1) lines would be collisionally deactivated (Ferland and Persson 1989); while densities N$_e\leq$10$^5$ implies extremely high electron temperatures (T$_e\geq$39000K) which are not consistent with the non-detection of any high energy transition. 

In summary, we are unable to constrain the physical condition of V1309 Sco line forming region without a detail model (which is not the scope of this paper), but it is reasonable to interpret the observations of  [OI] and [CaII] emission lines as evidence of a relatively dense chromosphere/envelope such to prevent the formation of other forbidden lines. 

Last, we also compared the intensity of the OI(1) multiplet ($\sim$7774~\AA) with that of the OI(4) triplet ($\sim$8446~\AA). Starting from epoch 4, the OI(4) intensity is significantly stronger than that of the OI(1) lines possibly implying fluorescence of the OI 1302 \AA \ ultra-violet line by the hydrogen Ly$\beta$. The Bowen fluorescence, or photo-excitation by accidental resonance (PAR), has been observed in a variety of astronomical objects and has been studied in classical novae by Kastner and Bathia (1995). Making use of their computations and plots and measuring our line intensities in epoch 4 and 5 spectra\footnote{Epoch 1 and 3 spectra do not show the 6300 [OI] line, while the OI(4) and (1) multiplets are partly in absorption and, therefore, difficult to measure reliably.}, we find only marginal agreement with the reported CNe observations and intensity ratios which are possibly consistent with quite low photo-excitation rates (considering the above density constraints). It has to be said, however, that Kastner and Bathia (1995) computations are made in the optically thin assumption and that, in optically thick conditions, all fluorescent lines are reduced in intensity.  
We conclude that PAR was certainly not present in the early epochs (1 to 3) of V1309~Sco decline and start possibly developing at the later epochs (from 4 on). It might very likely be present in the April SOAR spectra where we observe a relatively strong OI(4) emission lines, but not OI(1) (note that the wavelengths corresponding to the [OI](1) emission lines are not covered by the Goodman spectrograph).

\section{Discussion and conclusion}
Inspection of the AAVSO V-band data shows that V1309 Sco reached maximum light on Sep 6 2008 and declined with a relatively smooth light curve in the following 1.5 months. The same light curve shows that the nova $t_2$ time is $\sim$20 days, while its t$_3$ is about 25 days. However, the maximum of Sep 6, could have been a secondary one, the first having occurred on Sep 1 at visual magnitude $\sim$7.5 (see Fig.~\ref{fig0}) The scarce monitoring of the following days shows a sudden drop with a possible t$_2$ of only $\leq$3 days. 

The postion of V1309~Sco is nearly coincident with that of a red USNO-B1 star ($\sim$1$\sigma$) of magnitudes B=16.88 and R=14.80 (year of observation: 1966). Within $\leq$2$\sigma$ from the nova position there is a 2MASS object of J, H and K magnitudes equal to 13.282, 12.373 and 11.099, respectively. On a 1958 POSS-I~E/red plate,  the V1309~Sco progenitor has been identified with an object fainter than 19 mag (Jaques and Pimentel 2008, IAUC 8972; the mag limit of the POSS-I~E survey is 20). Depending on the correct progenitor, V1309~Sco outburst amplitude could either have been $\sim$7 mag or as large as 12 mags. 

The V1309~Sco light curve and spectral evolution described in the previous section are peculiar in the sense that the nova does not follow the prototype of any single class of mid-large outburst amplitude objects. Its  rapid spectral evolution toward a red continuum  with possible development of red-giant signatures after only a few weeks from the outburst (Rudy et al. 2008b, IAUC 8997) make it similar to the symbiotic recurrent novae  V745 Sco/89 and V3890 Sgr/90 (Williams et al. 1991). The postoutburst luminosities of the red continua for all three novae are many magnitudes greater than the preoutburst brightnesses, and require a photospheric radius that is orders of magnitude larger than that of a normal late-type giant, and substantially larger than the size of the Roche lobe of a CV with a period of order one day. 

However, both V745~Sco and V3890~Sgr showed a faster decline and a somewhat different spectral evolution. Sekiguchi et al. (1990) report t$_2$=5 and t$_3$=9 days for V745~Sco; while Anupama and Sethi (1994) measured t$_2$=12 and t$_3$=17 days, in the case of V3890~Sgr. 
Their outburst amplitude (A$\sim$7.3 mag for V745~Sco, Sekiguchi et al. 1990; 7$\leq$A$\leq$9 mag for V3890~Sgr, Wenzel 1990) are consistent with those typically observed in symbiotic recurrent novae such as RS-Oph and T~CrB. 

V745~Sco spectra were characterized by large velocities (FWHM$\sim$1000 km/s and extended wings up to 4000 km/s, Sekiguchi et al. 1990), and early development of high ionization energy emission lines (e.g. the 4640\AA \ blend and the HeII$\lambda$4686) and forbidden transitions ([OII](1), Sekiguchi et al. 1990, as well as [FeX], [FeVII] and [FeXI], Williams et al. 1991). Similarly Anupama and Sethi (1994) observations showed that in less than one month V3890~Sgr has developed forbidden coronal lines from [FeX], [FeXIV], [AX] and [AXI]. Rapid evolution and early development of high ionization potential emission lines and forbidden lines was also observed by Williams et al. (1991) within their survey and monitoring program for CN in outburst at CTIO. 

V1309 Sco spectra has not developed any high ionization coronal forbidden transition, yet, though the presence of the [CaII] doublet tends to be observed in those variables that display a late-type stellar continuum in the red, and therefore it might be the signature of emission from an extended chromosphere of the secondary star rather than ejecta from the surface of the white dwarf. 
The fact that the [CaII] $\lambda\lambda$7292,7324 lines are typically not observed in classical novae or nova-like variables is indicative of a density regime different from that of classical novae.

In addition, V1309 Sco velocities as measured from the emission lines FWHM  and their extended wings never exceeded 150 km/s and 1000 km/s, respectively\footnote{Higher velocities reported in the IAUCs reflect the lower spectral resolution of the instruments.}. 

We should further note that it is difficult to fit a symbiotic binary in the progenitor of V1309 Sco because of the missing giant companion. The NIR 2MASS colors measured for the nearby star mentioned above, dereddened using the measured E(B-V) and Cardelli et al. (1989) reddening law, provide J-H and K-H colors which are consistent with a M1 type giant (Frogel and Whitford 1987) at a distance of $\sim$11kpc. V1309~Sco is in the direction of the galactic center and hence, at a distance $<$8kpc. Though the two distance do not appear significantly different, we believe that V1309~Sco is much closer than 8~kpc because of the relatively small E(B-V) we have estimated from the maximum spectra. Hence, a cool giant companion in V1309~Sco progenitor should have brighter NIR magnitudes than those reported by 2MASS. 

It should be added that the presence of heavy element Fe-peak narrow absorption line systems in the early post outburst spectra is not completely unrelated to classical novae. 
These transient heavy element absorption systems have recently been observed in almost all novae studied at high spectral resolution (Williams et al. 2008), and their prominence in V1309~Sco is remarkably by far the most extensive such system observed so far. 

At the same time the red continuum, the narrow Balmer emission lines and the heavy-element absorptions are characteristic of the so called 'red-novae' such as V838~Mon/02 and the less well observed V4332 Sgr/94 and M31-RV/88. These objects all evolved in relatively short time toward M and K giant spectra, with V838~Mon developing the first L-giant spectrum ever claimed  (e.g. Munari et al. 2007). They have never shown evidence either of high ionization potential element emission lines nor they entered the nebular or coronal phases typically observed in CNe (e.g. Munari et al. 2007, Barsukova et al. 2007, Rushton et al. 2005, Banerjee  and Ashok 2002, Rich et al. 1989, Mould et al. 1990, Martini et al. 1999). Forbidden transitions from [OI] and [FeII] have been reported in the late spectra of V838~Mon ($>+$7 months since outburst, Wagner and Starrfield 2002, Munari et al. 2007, Kaminski et al. 2009) and V4332~Sgr ($\geq$+5 months since outburst, Martini et al. 1999). 
Large luminosities have been derived for M31-RV (M$_{bol}$=-10 mag, Rich et al. 1989, see also Mould et al. 1990) and V838~Mon (M$_V$=-9.8 mag, Sparks et al. 2008, see also Bond et al. 2003, Tylenda 2004 and Tylenda 2005 and reference therein, Soker and Tylenda 2003 and Tylenda et al. 2005); while the distance of V4332~Sgr is uncertain. In addition, the outburst light curve of the 3 objects differ in their time scale and (within the number of data points available for each of them) morphology, though Munari et al. (2007) have noted that they are ''remarkably similar'' once scaled by the time of their optical brightness free-fall. Munari et al. (2007) also noticed that all the three objects displayed the whole range of M type giants during such 4 mags free fall. We cannot do exactly the same comparison for V1309~Sco, but note that the presence of multiple/secondary maxima in the object light curve, makes it similar to V838~Mon (e.g. Goranskij et al. 2007).

V1309~Sco seems to share few peculiarities with symbiotic recurrent novae and more characteristics with the yet un-understood class of red-novae. Yet it cannot firmly be classified as belonging to this latter type of objects due to the lack of later epochs spectra and the possibly low luminosity. By placing V1309~Sco at the distance of 8 kpc and assuming m$_V$(max)=7.9 mag (see Fig~\ref{fig0}), we derive the upper limit of M$_V$=-8.3 mag, which is almost 2 mag fainter than the absolute magnitude derived for V838~Mon (Sparks et al. 2008) and M31-RV (Rich et al. 1989). However, should the red nova be explained by stellar merging phenomena, the maximum luminosity is not necessarily a stringent constraint. At the same time, V1309~Sco might represent a link (an intermediate case) between the two classes of symbiotic and red-novae, should the red-nova class be caused by a thermo-nuclear reaction on a small mass accreting white dwarf (Shara et al.2009, see also Iben and Tutukov 1992 and Idan et al. 2009 for a discussion about CNe outburst on small mass accreting white dwarfs) rather than to stellar merging (Tylenda and Soker 2006 and reference therein). 
The most recent models (Shara et al. 2009) for classical nova outburst on low mass white dwarf ($\sim$0.5M$_\odot$) accreting at a low rate (a few 10$^{-11}$\.M/yr) predict that these type of novae will accumulate large amount of mass on the white dwarf surface, before the TNR ignition. The outburst will result in  massive ($\sim$10$^{-3}$M$_\odot$) cool, red ejected envelopes. In addition multiple maxima, tremendous absolute magnitudes (up to M$_V\sim$-9$\div$-13 mag or luminosity L$\sim$10$^6$L$_\odot$), low expansion velocities and oxygen rich and shocked spectra are predicted too, thus fitting the main characteristics  observed in the red-novae.

However, recent observation in high resolution spectroscopy of V838 Mon (Kaminski et al. 2009) favor the young planet merging in a triple/multiple system within a open cluster. Hence, whether V838 Mon could be explained by the CN outburst on a small mass WD remains uncertain, doubtful and highly debated. In addition, whether it is the prototype of the red novae variables or just a peculiar object has to be established as well. 

A class of novae  hosting a small mass white dwarf should exist and may already have been observed.   
Whether V1309~Sco belong to such a subclass or instead is a red-nova fitting the star-merging scenario can only be established through further spectroscopic observations. Optical and NIR spectroscopy should enable the identification of 1) the late epoch evolution of the object, possibly toward later giant spectral types, 2) the development of high excitation coronal lines and forbidden transitions, 3) the possible presence of a blue companion 4) radial velocity shifts which could be ascribed to orbital motion. In the case of a binary symbiotic-like system, the orbital periods are expected to be of the order of a few hundreds of days. Significantly shorter orbital periods (a few hours to day) would imply a dwarf donor companion.

\begin{acknowledgements}
We thank the ESO director general T. De Zeeuw for having
allocated Director Discretionary Time at the UT2+UVES allowing the collection
of data published in this paper. We acknowledge with thanks the variable star observations from the AAVSO International Database contributed by observers worldwide and used in this research. We also thank the referee Rushton M.T. for the careful reading of the manuscript and the useful and detailed report. 
This work was finished in Monte Porzio at the Rome Observatory which kindly hosted EM for 1 month science leave. 
\end{acknowledgements}

\end{document}